%% file: decode20030717.tex
\documentclass[aps,pra,nofootinbib,groupedaddress,amsfonts,twosided,twocolumn,amssymb,final]{revtex4}
\usepackage{amsthm}
\usepackage{amsmath}
\usepackage{amsfonts}
\usepackage{amssymb}
\usepackage{epsf}
\usepackage{revsymb}
\usepackage[dvips]{color}
\usepackage{amscd} 
\def\1#1{{\bf #1}}
\def\2#1{{\cal #1}}
\def\4#1{{\tt #1}}
\def\5#1{{\sf #1}}
\def\6#1{{\mathfrak #1}}
\def\7#1{{\Bbb #1}}
\def\8#1{{\rm #1}}
\def\9#1{{\mathcurl #1}}

\newtheorem{thm}{Theorem}[section]

\newtheorem{lem}{Lemma}[section]

\newtheorem{cor}[thm]{Corollary}

\def\<{\langle}
\def\>{\rangle}

\def\H#1{\9L_2(\7F^{#1})}

\DeclareFontFamily{OT1}{rsfs}{}
\DeclareFontShape{OT1}{rsfs}{m}{n}{ <-7> rsfs5 <7-10> rsfs7 <10-> rsfs10}{}
\DeclareMathAlphabet\mathcurl{OT1}{rsfs}{m}{n}
\begin{document}
%%%
\title{Quantum error correcting codes and one-way quantum computing:\\
Towards a quantum memory}
\author{Dirk~Schlingemann}
\affiliation{Institut f{\"u}r Mathematische Physik, TU-Braunschweig
\\ 
Mendelssohnstra{\ss}e 3
\\
38106 Braunschweig, Germany}
%%%
\begin{abstract}
For realizing a quantum memory we suggest to first encode quantum information via a quantum error correcting code and then concatenate combined decoding and re-encoding operations. This requires that the encoding and the decoding operation can be performed faster than the typical decoherence time of the underlying system. The computational model underlying the one-way quantum computer, which has been introduced by Hans Briegel and Robert Raussendorf, provides a suitable concept for a fast implementation of quantum error correcting codes. It is shown explicitly in this article is how encoding and decoding operations for stabilizer codes can be realized on a one-way quantum computer. This is based on the graph code representation for stabilizer codes, on the one hand, and the relation between cluster states and graph codes, on the other hand.
\end{abstract}
%%%
\maketitle
%%%
\section{Introduction}
%%%
The concept of quantum error correcting codes plays a central role for the realization of quantum computational processes. In particular, quantum information, that is stored in a quantum system, has to be protected against decoherence. The states of the ``input system" describe the quantum information we wish to store. These input states are given by density operators $\rho\in\9L_1(\2K)$ acting on an ``input Hilbert space" $\2K$. The second system under consideration is the ``output system" whose states are the density operators on an ``output Hilbert space" $\2H$. 

The output system is the one which is present in nature, and in which we wish to encode quantum information. All relevant decoherence processes operating on the output system due to the coupling with the environment. A possible description of decoherence is the following: Consider a one parameter semi-group $t\mapsto T_t$, $T_t T_s=T_{t+s}$, of channels\footnote{In the Heisenberg picture, a channel $T$ is a completely positive map that preserves unit operator.} acting on the observable algebra of all linear operators on $\2H$. The channel $T_t$ is interpreted as the process of decoherence (error) that is present at the ``time" $t$. A quantum computational process is only sufficiently reliable if the effect of errors is below a threshold $\epsilon$. We define the ``decoherence time" of the system to be the largest time $s$ for which the cb-norm $\|T_s-\8{id}\|\leq \epsilon$ is below the tolerable threshold. For most of the systems, which can be realized in experiments, the decoherence time $s$ is too small for a sensible quantum memory.

In order to protect quantum information, we encode quantum states of the input system into quantum states of the output system. In the Heisenberg picture, which is preferably used here, an encoding operation is described by a channel $E$ that maps the observable algebra $\9B(\2H)$ of the output system into the observable algebra $\9B(\2K)$ of the input system. For receiving the encoded quantum information back, we also need a decoding operation which is a channel $D$ that maps the input observable algebra into the output observable algebra. 

We encode states of the input system via a channel $E$ to realize a quantum memory. After a certain time $t$, the encoded quantum information is corrupted due to the decoherence process $T_t$. The stored information is recovered by an application of an appropriate decoding operation $D$. The total channel that has been performed operates on the input system and it is given by the composition $ET_tD$. If this channel is close to the identity $\|ET_tD-\8{id}\|\leq\epsilon$, we would have stored our quantum information successfully during the time $t$. For a reasonable coding scheme, the storing time $t$ is much larger than the decoherence time $s$ of the output system. The ratio $t/s$ can be increased by increasing the ratio $\8{dim}(\2H)/\8{dim}(\2K)$ of the dimension of the output system and the dimension of the input system.
Thus, increasing the storing time requires a larger amount of resources. 

An alternative idea for obtaining large storing times is to concatenate decoding and re-encoding operations. Suppose we are able to store quantum information successfully for a time $t$, i.e the channel  
$ET_tD$ is close to the identity. Then we just recode the corrected quantum information again, which corresponds to the operation $ET_tDE$. Again the system undergoes decoherence for a further time $t$. Then the decoding operation is applied once more. The channel $ET_tDET_tD$ is close to the identity and we have stored our quantum information successfully for a time period $2t$. This heuristic picture is only realistic, if we assume that the decoding re-endoding operation $DE$ can be performed much faster than the typical decoherence time $s$ of the output system.   

For realizing a quantum memory, we are therefore interested in ``fast implementations" for encoding and decoding operations. The model of one-way quantum computing, introduced by Robert Raussendorf and Hans Briegel \cite{BrieRau00,BrieRau01,BrieRau01b,BrieRau01c,BrieBrowRau02,BrieRau02,BrieRauSchenz02,BrieBrowRau03} suggest to be an appropriate base for realizing fast operations due to the intrinsic parallelism of this model. A one-way quantum computer operates on a system of qudits (quantum digits). Elementary operations within this scheme are 
\begin{itemize}
\item
local preparation procedures which address every qudit individually,      
\item
one elementary step of an dynamics that corresponds to two-qudit interactions,  
\item 
local measurement operations which operate independent on each qudit,
\item
and conditional local unitary operations (depending on the measurement outcomes). 
\end{itemize}
The parallelism of a one-way quantum computer is based on the dynamics which is a global operation that addresses all qudits at the same time. 

The pattern of two-qudit interactions defines a ``weighted graph" on the set of qudit positions. Namely, two positions are connected by an edge, if the corresponding qudits interact with each other. The ``weight", attachted to an edge, is the ``strength" of the qudit coupling (which is an integral number for suitable interactions).   

As we have discussed in previous articles \cite{SchlWer00,Schl02}, an encoding operation, called ``graph code", can be associated a ``weighted graph" in a natural manner and there is in fact a close relation between one-way quantum computing and error correction via graph codes. The qudits under consideration are grouped into ``input" qudits and ``output qudits". In \cite{Schl03} we have shown that the following is true:   

{\em Every encoding procedure for a graph code can be implemented on a one-way quantum computer by four elementary operations:} (1) First one applies an appropriate local preparation procedure: Every output qudit is prepared in the ``standard state" $\frac{1}{\sqrt{d}}(|0\>+|1\>+\cdots +|d-1\>)$. (2) One elementary step of a discrete dynamics is performed. This dynamics corresponds to the interaction pattern which is given by the underlying graph. (3) The input qudits are measured independently in the ``$x$-basis", a suitable generalization of the $\sigma_x$-eigenbasis to the non-binary case. (4) Depending on the measurement outcome, a suitable local unitary operation is performed.
 
This is a scheme can be applied to all stabilizer codes \cite{Got97}, since every stabilizer code has a graph code representation as it is shown in \cite{Schl02,GraKlaRoe02}.
 
The main result of the paper is concerned with an implementation of the decoding operation of a graph code. It is based on a suitable extension of the coding graph by adding ``syndrome vertices" and edges that connect the syndrome vertices with the output vertices in an appropriate manner. The syndrome vertices are the positions of the syndrome qudits which are used to measure the ``error syndrome". 

{\em Every decoding operation can also be implemented one a one-way quantum computer by a sequence of four elementary operations:} (1) A local preparation procedure is performed: The input qudits as well as the syndrome qudits are prepared in the ``standard state". (2) One inverse elementary step of the discrete dynamics, which corresponds to the extended coding graph, is performed. (3) The output qudits are measured independently in the $x$-basis and the syndrome qudits are measured independently in the $z$-basis which generalizes the $\sigma_z$-eigenbasis to the non-binary case. (4) Depending on the measurement outcome of the output and syndrome degrees of freedom, a  suitable local unitary operation is performed.

We mention at this point, that both operations, the encoding and the decoding, are based on the same dynamics. This is of course what one expects as far as the implementation of stabilizer codes by quantum circuits (consisting of one- and two-qudit elementary gates) is concerned \cite{Schl02b}. For the decoding one just uses the reversed circuit.   
 
The paper is organized as follows: In Section~\ref{sec-1} we discuss the structure for encoding and decoding operations in view of the stabilizer formalism for deriving an explicit expression for the decoding operation. 

Section~\ref{stab-codes} reviews the concept of stabilizer codes and their graph code representations. It is also shown here how to construct explicitly an error correcting scheme from a given graph being associated with a quantum error correcting code.

The content of Section~\ref{impl-1} begins with a brief description of the concept of one-way quantum computing. Then we present here, with help of the results derived in the previous sections, the an implementation for encoding and decoding operations underlying the model of the one-way quantum computer.

%%%
\section{On the structure of encoding and decoding operations}
%%%
\label{sec-1}
Within this section, we describe the structure for encoding and decoding operations in view of the stabilizer formalism. We denote by $\2K$ the Hilbert space for the ``input system". The density operators on $\2K$ represent input states which carry the information we wish to protect against decoherence. The states of the ``output system", in which the information is encoded, are given by density operators on a larger Hilbert space $\2H$ with $\8{dim}\2H>\8{dim}\2K$. The effects of decoherence are described (in the Heisenberg picture) by a channels\footnote{Completely positive unit-preserving maps on $\9B(\2H)$} on the observable algebra $\9B(\2H)$ of the output system. 

%%%
\subsection{Error correcting schemes}
%%%
\label{sec-1-1}
As ``error basis", we consider a family of $\6w=(\1w_x|x\in X)$ of linearly independent unitary operators on the output Hilbert space $\2H$. The effects of decoherence are described by the set of all channels $\9T$ on $\9B(\2H)$ whose Kraus operators are linear combinations of basis elements in $\6w$. Thus, each channel $T\in\9T$ has the form
\begin{equation}
T(a)=\sum_{x,y\in X} t_{x,y} \1w_x^*a\1w_y
\end{equation}      
where $t=(t_{x,y}|x,y\in X)$ is a positive matrix. 

A {\em complete family of mutually orthogonal isometries} $\6v=(\1v_g|g\in G)$ consists of isometries $\1v_g$ from the input Hilbert space $\2K$ into the output Hilbert space $\2H$ such that the ranges of $\1v_g$ and $\1v_h$ are orthogonal if $g\not=h$, i.e. $\1v_g^*\1v_h=\delta_{g,h}\11_\2K$ 
and each vector in $\2H$ is contained in the range of one isometry $\1v_g$, i.e. $\sum\1v_g\1v_g^*=\11_\2H$.\footnote{The set $G$ has finite cardinality $|G|=\frac{\8{dim}\2H}{\8{dim}\2K}$. A necessary and sufficient condition for the the existence of a complete family of mutually orthogonal isometries is that the dimension of input Hilbert space divides the dimension of the output Hilbert space.} 

The elements of $G$ can be interpreted as error syndromes, whereas the elements in $X$ corresponds to the different types of errors. An error correcting scheme relates an error $x\in X$ to a syndrome $g\in G$. This relation tells us what correction operation $h\in H$ has to be performed, where $H$ is a finite set whose elements correspond to the possible correction procedures. 

\paragraph*{Definition.}
An {\em error correcting scheme} is a four-tuple $(\6w,\6v,\6u,\gamma)$ which consists of a unitary error basis $\6w=(\1w_x|x\in X)$ in $\9B(\2H)$, a complete family $\6v=(\1v_g|g\in G)$ of mutually orthogonal isometries, a family $\6u=(u_h|h\in H)$ of unitary operators on the input Hilbert space $\2K$, and a function $\gamma\mathpunct:X\times G\times H\to \{0,1\}$ such that the following conditions are fulfilled:
\begin{enumerate}
\item
For each triple $(x,g,h)\in X\times G\times H$ the relation $\gamma(x,g,h)=0$ implies that  
\begin{equation}\label{ec-scheme}
\1w_x\1v_e=\1v_{g}\1u_{h}
\end{equation}
holds for some $e\in G$.
\item
For each $x\in X$ there exists a pair $(g,h)\in G\times H$ such that $\gamma(x,g,h)=0$.
\item
Suppose $\gamma(x,g,h)=0$ and $\gamma(x',g,h')=0$ holds for $x,x'\in X$, $h,h'\in H$ and $g\in G$, then $h=h'$ follows. 
\end{enumerate}

The constraint $\gamma(x,g,h)=0$ relates an error syndrome $g\in G$ to an error $x\in X$ and an appropriate  correction procedure $h\in H$. Equation (\ref{ec-scheme}) describes the fact that the chosen scheme $h$ indeed eliminates the error $x$ that occurred. The function $\gamma$ that fulfill the conditions 2. and 3. is called a ``syndrome table".

%%%
\subsection{Quantum error correcting codes associated with error correcting schemes}
%%%
\label{sec-1-2}
To each error correcting scheme $(\6w,\6v,\6u,\gamma)$ we {\em associate a quantum error correcting code $(E,D)$} which consists of an encoding and a decoding channel in order to correct the errors which are caused by noisy channels in $\9T$. 

The {\em encoding} is the channel $E$ that is implemented by a single isometric Kraus operator $\1v_e$. That is, an output observable $a\in\9B(\2H)$ is mapped to the input observable $E(a)=\1v_e^* a\1v_e$. An appropriate {\em decoding channel} $D:=SC$ can be composed of a syndrome measurement $S$ and a correction operation $C$.  

\paragraph*{Syndrome measurement.}  The syndrome measurement is an operation which assigns to a state of the output system a state of the input system together with a classical measurement result, the error syndrome $g\in G$. In the Heisenberg picture, the syndrome measurement is a channel $S$ that maps an operator valued function $b\mathpunct:G\to\9B(\2K)$ to the operator  
\begin{equation}\label{kraus-syndrome}
S(b)=\sum_{g\in G} \1v_g b(g)\1v_g^*
\end{equation}
of the output observable algebra $\9B(\2H)$. 

\paragraph*{Correction procedure.} The correction procedure is a conditional unitary operation which is performed on the input system. Depending on the outcome $g\in G$ of the syndrome measurement, it transforms a state of the input system. In the Heisenberg picture, the correction is described by a channel $C$ which maps an observable $b$ of the input system to an operator valued function $C(a)\mathpunct:G\to\9B(\2K)$ which is given by  
\begin{equation}\label{corr-1}
C(b)(g):=\sum_{(x,h)} c(x,g,h) \1u_h b \1u_h^* \; .
\end{equation}
We have introduced here the positive function $c$ on $X\times G\times H$ according to   
\begin{equation}
c(x,g,h)=\left\{\begin{array}{ccc}
n(g,h)^{-1}&\text{if}&\gamma(x,g,h)=0\\
|X|^{-1}\delta_{\iota,h}&\text{if}&g\in G_\gamma\\
0&\text{else}&
\end{array}\right. 
\end{equation}
where $\iota$ is some element in $H$. Moreover, $n(g,h)$ is the number of error labels $x\in X$ which solves $\gamma(x,g,h)=0$ for a given error syndrome $g$ and a correction procedure $h$ and $G_\gamma$ is the set of ``left-over-syndromes" $g$ for which there there exists no pair $(x,h)\in X\times H$ such that $\gamma(x,g,h)=0$.

The function $c$ can by viewed as a ``classical analysis" of the received syndromes: The measurement result $g$ is feeded into a program that checks whether an error $x$ and a correction operation $h$ is related to $g$ by $\gamma(x,g,h)=0$. 

One observes from the properties of the syndrome table $\gamma$ that $C$ is an algebra homomorphism. Namely, for each $g\in G\setminus G_\gamma$ there exists a unique $h(g)\in H$ such that $\gamma(x,g,h(g))=0$ holds for some $x\in X$. This implies that the identity   
\begin{equation}\label{corr-2}
C(b)(g)=\left\{\begin{array}{ccc} \1u_{h(g)}b\1u_{h(g)}^*&\text{if}&g\in G\setminus G_\gamma\\
\1u_\iota b\1u_\iota^*&\text{if}&g\in G_\gamma
\end{array}\right.
\end{equation}
is valid for each observable $b$ of the input system.

We show in Appendix~\ref{app-1} that each noisy channel $T\in\9T$ is indeed completely corrected by the code $(E,D)$: 
%%%
\begin{thm}\label{thm-1}
Let $(\6w,\6v,\6u,\gamma)$ be an error correcting scheme and let $(E,D)$ be the error correcting code associated with it. Then for all $T\in\9T$ the identity
\begin{equation}
ETD=\8{id}_{\9B(\2K)}
\end{equation}
is valid. In particular, the decoder $D$ is a *-algebra homomorphism\footnote{A *-algebra homomorphism is a $D$ is a linear map that preserves the multiplicative structure, $D(b_1b_2)=D(b_1)D(b_2)$ and the adjoint $D(b)^*=D(b^*)$.}.
\end{thm}
%%%

%%%
\section{Error correcting schemes for stabilizer codes} 
\label{stab-codes}
In the first part of this section, we give a brief review on the graph code representation of stabilizer codes. The second part deals with an explicit construction of error correcting schemes from a given graph code.

%%%
\subsection{Stabilizer codes}
%%%
\label{stab-codes-1}
The classical configuration space of a ``digit" is given by a finite ``alphabet" which is a finite field $\7F$ of order $d$\footnote{For binary systems we are concerned with the field of two elements $\7F_2=\{0,1\}$.}. A classical register is described by its configurations which are given by tuples $q^I=(q^i|i\in I)$ in the vector space $\7F^I$ over $\7F$. Each position $i\in I$ of the register accepts a letter $q^i$ from the alphabet $\7F$. The corresponding {\em phase space} of a register is modeled by the vector space $\Xi^I=\7F^I\oplus\7F^I$. The first vector entry of a point $\xi^I=(p^I,q^I)\in\Xi^I$ is interpreted as {\em momentum} and the second entry as the {\em position}.

The Hilbert space, describing a quantum register of qudits, is the space $\H{I}$ of complex valued functions on $\7F^I$ and its complex dimension is $d^{|I|}$ where $|I|$ is the number of elements in $I$. The scalar product of two functions $\psi_1,\psi_2$ is given by
\begin{equation}
\langle\psi_1,\psi_2\rangle=\int\8dq^I\ \bar\psi_1(q^I)\psi_2(q^I)
\end{equation}
where the integration is performed with respect to the ``normalized Haar measure" of the additive group $\7F^I$. 

The {\em shift operator} $\1x(q^I)$, associated with position vector $q^I$, is the unitary operator, which translates a function $\psi\in\H{I}$ by $q^I$
\begin{equation}
(\1x(q^I)\psi)(q_1^I)=\psi(q_1^I-q^I)
\end{equation}
and the {\em multiplier operator} $\1z(p^I)$, which is associated with the momentum $p^I$ is the unitary multiplication operator, acting on a function $\psi\in\H{I}$ by
\begin{equation}
(\1z(p^I)\psi)(q^I)=\chi(p^I,q^I)\psi(q^I) \; .
\end{equation}
The phases $\chi(p^I,q^I)$ form a symmetric bicharacter \footnote{Let $\varepsilon\mathpunct:\7F\to \8U(1)$ is a faithful character of the additive group $\7F$. Then an appropriate bicharacter $\chi$ is defined by $\chi(p^I,q^I)=\prod_i\varepsilon(p^i q^i)$.} of the additive group $\7F^I$ \cite{SchlWer00}.
 
This allows to assign to each point in phase space $\xi^I=(p^I,q^I)\in\Xi^I$ the unitary ``Weyl" operator
$\1w(\xi^I):=\1z(p^I)\1x(q^I)$. The Weyl operators satisfy a discrete version of the canonical commutation relations
\begin{equation}\label{discr-ccr}
\1w(\xi_1^I)\1w(\xi_2^I)=\chi(p_2^I,q_1^I)\1w(\xi_1^I+\xi_2^I)\; .
\end{equation}
They form a basis for the algebra $\6A(I)$ of all linear operators on the Hilbert space $\H{I}$. For a subset $J\subset I$ of positions, the Weyl operators $\1w(\xi^J)$ generate sub-algebra $\6A(J)\subset\6A(I)$, identifying a quantum sub-register by operating only non-trivially on the positions $J$. 

According to the Weyl commutation relations we obtain an abelian sub-algebra  $\6A(\Lambda)\subset\6A(J)$, the ``stabilizer algebra", which is generated by the set of Weyl operators\footnote{The following notation is used: For a matrix $\Theta^N_M$ and for two subsets $K\subset M$, $L\subset N$, we write $\Theta^L_K=(\Theta(l,k)|l\in L,k\in K)$ for the corresponding sub-block.}
$\1w(\Lambda^{J}_{J}q^{J},q^{J})$, $q^{J}\in\8{ker}\Lambda^{I}_{J}$. The stabilizer algebra only depends on the subgraph $\Lambda^{IJ}_{IJ}$ corresponding to the removing of syndrome vertices. However, the edges that connect syndrome vertices with others play an important role for the representation theory of the stabilizer algebra and we are concerned with graphs that fulfill the following list of conditions \cite{Schl02,Schl03}:

\paragraph*{Definition.} A weighted graph on the union of input vertices $I$, output vertices $J$, and syndrome vertices $L$ is called {\em admissible} if its adjacency matrix $\Lambda$ fulfills the following conditions:  
\begin{enumerate}
\item
The blockmatrix $\Lambda^{J}_{IL}$ is invertible with an inverse $\bar\Lambda^{IL}_J$.

\item There are no edges that connect input and syndrome vertices, i.e. the block matrix $\Lambda^{IL}_{IL}=0$ vanishes.
\end{enumerate}

The figure (FIG.~\ref{example}), given below, shows two simple examples for admissible graphs.
%%%
\begin{figure}[h]
\begin{center}
\epsfxsize=8.7cm\epsffile{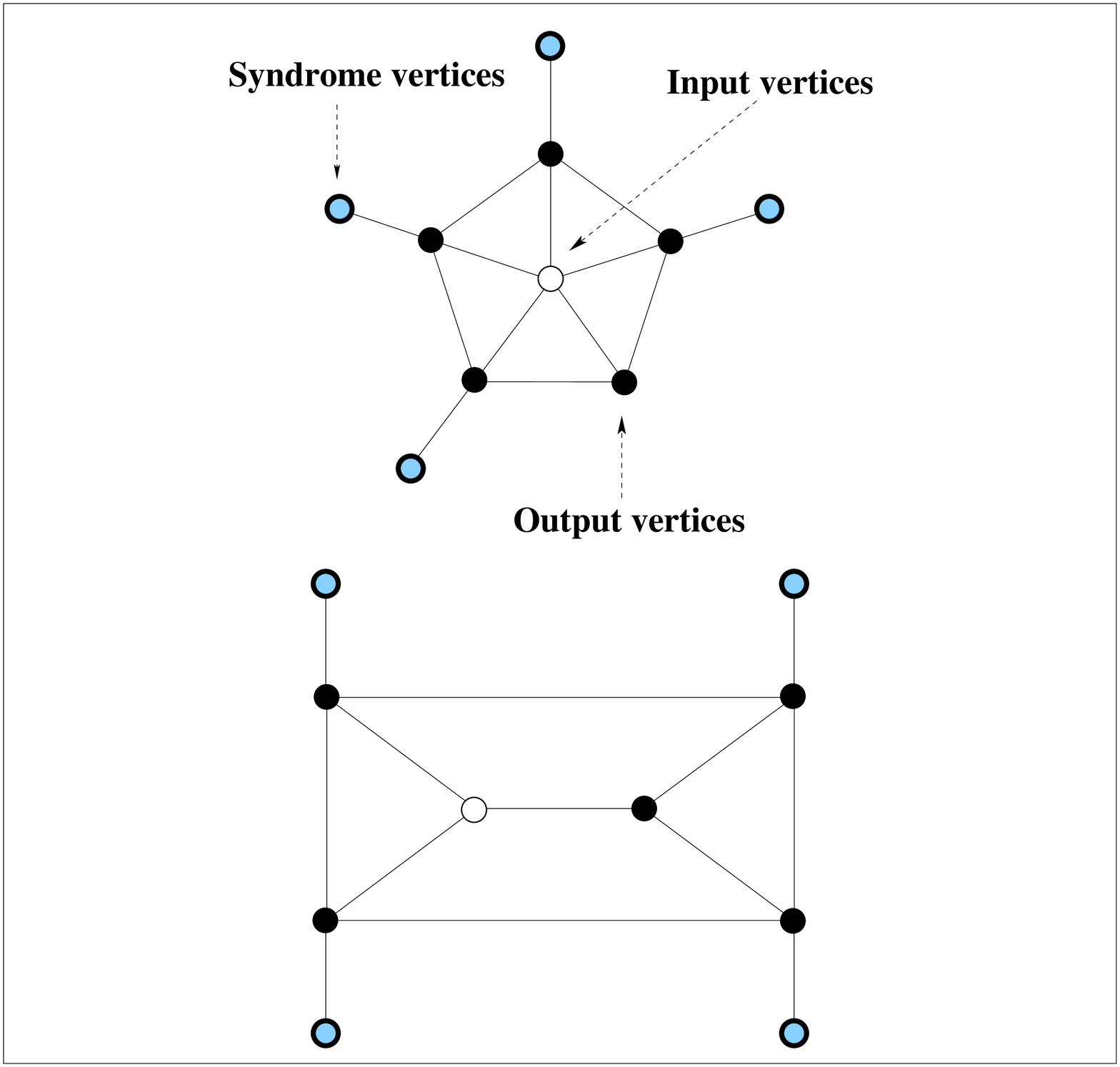}
\end{center}
\caption{Examples for admissible graphs that correspond to $[[5,1,3]]_d$ stabilizer codes.}
\label{example}
\end{figure}
%%% 

Since $\6A(\Lambda)$ is an abelian C*-algebra its representation on $\H{J}$ can be decomposed into irreducible representations. From an admissible graph $\Lambda$ we construct for each $q^L\in\7F^L$ an isometry $\1v_{[\Lambda,q^L]}\mathpunct:\H{I}\to\H{J}$ which is defined by operating on a vector $\psi\in\H{I}$ according to: 
\begin{equation}\label{graph-code}
(\1v_{[\Lambda,-q^L]}\psi)(q^J)=\sqrt{d}^{|I|}\int\8dq^{I} \ \tau(\Lambda,q^{IJL})\psi(q^I) \; .
\end{equation}
The phases $\tau(\Lambda,q)$, $q\in\7F^{IJL}$, fulfill the relation of a one-dimensional projective representation: $\tau(\Lambda,q_1+q_2)=\tau(\Lambda,q_1)\tau(\Lambda,q_2)\chi(\Lambda q_1,q_2)$.

Employing the results of \cite{Schl03}, the vectors in the range of $\1v_{[\Lambda,q^L]}$ are joint eigenvectors for the stabilizer algebra. In fact the identity 
\begin{equation}
\1w(\Lambda^J_{J}q^J,q^J)\1v_{[\Lambda,q^L]}\psi=\tau(\Lambda,q^{JL})\1v_{[\Lambda,q^L]}\psi 
\end{equation}
holds for all vectors $\psi$ in the input Hilbert space $\H{I}$ if $q^J$ fulfills the identity $\Lambda^{I}_{J}q^{J}=0$. 

The Kraus operators of channels, that cause decoherence on a certain number $t$ of qudits, are linear combinations of Weyl operators $\1w(\xi^J)$ for which the phase space vector $\xi^J$ has ``weight" less or equal to $t$. The weight of a vector $\xi^J=(\xi^j|j\in J)$ is the number of non-zero components $\xi^j\not=0$. We denote by $\Xi^J_t$ the subset of all phase space vectors with weight less or equal to $t$.

We say that an admissible graph $\Lambda$ is associated with a $t$-error correcting code if the Knill-Laflamme condition \cite{KnLafl95}
\begin{equation}
\1v_{[\Lambda,0^L]}^*\1w(\xi_1^J)^*\1w(\xi_2^J)\1v_{[\Lambda,0^L]}\in\7C\11_I
\end{equation}
holds for all $\xi_1^J,\xi_2^J\in\Xi_t^J$. We recall the result of \cite{SchlWer00}: The error correcting code associated with $\Lambda$ is a $t$-error correcting code, if and only if 
\begin{equation}
\Lambda^{J\setminus E}_{IE}q^{IE}=0 \; \text{ implies } \; q^{I}=0 \; \text{ and } \; \Lambda^I_Eq^E=0 
\end{equation}
is true for all sets $E$ that contain at most $2t$ elements.

%%%
\subsection{Error correcting schemes for $t$-error correcting codes}
%%%
\label{stab-codes-2}
An error-correcting scheme can be constructed from an admissible graph $\Lambda$ for a $t$-error correcting code. It consists of a unitary error basis $\6w_\Lambda$, a complete set of mutually orthogonal isometries $\6v_\Lambda$, a family of unitary operators $\6u_\Lambda$, which implement the correction operation, and a syndrome table $\gamma_\Lambda$ which relates the error syndrome to the correction operation. We construct each of the four objects sequentially:

\begin{enumerate}
\item 
First, we choose a convenient unitary error basis $\6w_\Lambda$ which consists of multiples $\1w_{[\Lambda,\xi^J]}:=\tau(\Lambda,q^J)\1w(\xi^J)$ of Weyl operators for which the phase space vector $\xi^J=(p^J,q^J)\in\Xi^J_t$ has weight less or equal to $t$. 

\item 
According to the result of \cite{Schl03}, the set $\6v_\Lambda$, which consists of the isometries $\1v_{[\Lambda,q^L]}$, $q^L\in\7F^L$, form a complete set of mutually orthogonal isometries.

\item 
For the correction operation, we just can take all the Weyl operators $\6u_\Lambda$ that act on the input Hilbert space $\H{I}$. In other words, $\6u_\Lambda$ consists of the Weyl operators  $\1u_{[\Lambda,\xi^I]}=\1w(\xi^I)$ with  $\xi^I\in\Xi^I$.

\item 
As a candidate for a syndrome table, we take the function  $\gamma_\Lambda\mathpunct:\Xi_t^J\times\7F^L\times\Xi^I\to \{0,1\}$ which is defined as follows: If the relations 
\begin{equation}\label{gam}
\begin{array}{lcl}
p^J-\Lambda^J_{IJL}q^{IJL}&=&0\\
&&\\
p^I-\Lambda^I_{J}q^{J}&=&0
\end{array}
\end{equation} 
are fulfilled then we put $\gamma_\Lambda(\xi^J,q^L,\xi^I)=0$. In all other cases, we set $\gamma_\Lambda(\xi^J,q^L,\xi^I)=1$. 
\end{enumerate}

In fact, we obtain a error correcting scheme by this construction procedure as we prove in Appendix~\ref{app-2}: 
%%%
\begin{thm}\label{e-c-scheme}
Let $\Lambda$ be an admissible graph associated with a $t$-error correcting code. Then the four-tuple $(\6w_\Lambda,\6v_\Lambda,\6u_\Lambda,\gamma_\Lambda)$, given above, is an error correcting scheme.
\end{thm}
%%%

%%%
\section{Implementing error correcting schemes by a one-way quantum computer}
%%%
\label{impl-1}
By making use of the results of Section~\ref{sec-1}, we obtain an explicit expression for the encoding and decoding operation $(E_\Lambda,D_\Lambda)$ that is associated with the error correcting scheme $(\6w_\Lambda,\6v_\Lambda,\6u_\Lambda,\gamma_\Lambda)$.

After a brief introduction into the basic elementary operations of one-way quantum computing, we apply the results of \cite{Schl03} to derive an implementation of both, the encoding and decoding operation, on a one-way quantum computer.

%%%
\subsection{One-way quantum computing}
%%%
\label{one-way}
We now introduce a class of operations which are viewed as elementary concerning the concept of one-way quantum computing. For this purpose it is convenient to fix one normalized standard vector $\Omega_K\in\H{K}$ 
which we choose to be the constant function on $\7F^K$. By applying different multiplier operators, this yields an orthonormal basis which consists of the vectors $\1z(p^K)\Omega_{K}$, $p^K\in\7F^K$. This basis is called the $x$-basis, which is the joint eigenbasis of the shift operators. Any other product basis can by obtained by applying a local unitary operator $U_K=\otimes_k U_k$ to the $x$-basis. In particular, the local Fourier transform $F_K$, which is given by
\begin{equation}
(F_K\psi)(p^K)=\sqrt{d}^{|K|}\int\8dq^K \chi(p^K,q^K)\psi(q^K)
\end{equation}
maps transforms the $x$-basis to the so called $z$-basis which is nothing else but the joint 
eigenbasis of the multiplier operators.

%%%
\paragraph*{Local preparation.}
To each local unitary operator $U_I=\otimes_{i\in I} U_i$ we associate the channel $E_{[U_I]}$ which maps an operator $a\in\6A(IJ)$ to the operator
\begin{equation}
E_{[U_I]}=\Phi_I^*U_I^* a U_I\Phi_I \in\6A(J) \; .
\end{equation}
Here $\Phi_I$ is the isometry which assigns to a vector $\psi$ the tensor product $\Phi_I\psi=\psi\otimes\Omega_I$. The channel $E_{[U_I]}$ describes the {\em local preparation} of each individual qudit at position $i\in I$ in the state $U_i\Phi_i$.  

%%%
\paragraph*{Local measurements.} 
Local measurements are dual to the local preparation schemes. We associate now to each local unitary operator $U_I$ a channel $M_{[U_I]}$ which maps an operator valued function $a\mathpunct:\7F^I\to\6A(J)$ 
to the operator 
\begin{equation}
M_{[U_I]}(a)
=\sum_{p^I\in\7F^I} U_I\1z(p^I)\Phi_I \ a(p^I) \ \Phi_I^*\1z(p^I)^*U_I^* 
\end{equation}
in $\6A(IJ)$. The operators $U_I\1z(p^I)\Phi_I$, $p^I\in\7F^I$, are a complete family of mutually orthogonal isometries. This implies, in particular, that $M_{[U_I]}$ is an algebra homomorphism. Concerning the Heisenberg picture, this is the characteristic property of a projection valued measure.

For our purpose, there are two intersting measurement bases: The $x$-basis corresponding to $U_I=\11_I$ and the $z$-basis corresponding to $U_I=F_I$. 

%%%
\paragraph*{Elementary step of a discrete dynamics.} 
Let $\Gamma$ be the adjacency matrix of a weighted graph with vertices $K$. We define the unitary multiplication operator $u(\Gamma)$ according to 
\begin{equation}
(u(\Gamma)\psi)(q^K):=\tau(\Gamma,q^K)\psi(q^K)
\end{equation}
which implements an automorphism $\alpha_{[\Gamma]}$ of $\6A(K)$ by mapping an observable $a$ to $\alpha_{[\Gamma]}(a)= u(\Gamma)^*a u(\Gamma)$.

%%%
\paragraph*{Conditional phase space translations.} 
A classical device which maps a probability distribution on the register configurations $\7F^L$ to a probability distribution on the phase space $\Xi^K$ is described, in the Heisenberg picture, by a channel $A$ which maps the classical observable algebra of functions $\9C(\Xi^K)$ on phase space $\Xi^K$ to the algebra $\9C(\7F^L)$ of functions on the register configurations $\7F^L$.    

To a classical channel $A$ we associate the {\em conditional phase space translation}  $C_{[A]}$ which is a channel that assigns to an observable $a\in\6A(K)$ an operator valued function $C_{[A]}(a)\mathpunct:\7F^L\to \6A(K)$ according to 
\begin{equation}
C_{[A]}(a)(q^L)=\sum_{\xi^K\in\Xi^K} A(q^L,\xi^K) \ \1w(\xi^K)a\1w(\xi^K)^* 
\end{equation}
where the classical channel $A$ acts on a function $f\in\9C(\Xi^K)$ by $(Af)(q^L)=\sum_{\xi^K} A(q^L,\xi^K)f(\xi^K)$. 

%%%
\subsection{The encoding operation} 
%%%
\label{impl-1-2}
The encoding operation, associated with the error correcting scheme $(\6w_\Lambda,\6v_\Lambda,\6u_\Lambda,\gamma_\Lambda)$, only depends on the subgraph $\Lambda^{IJ}_{IJ}$ of the admissible graph $\Lambda$ on the union $IJ$ of input and output vertices. The encoding is implemented by the single isometric Kraus operator $\1v_{[\Lambda,0^L]}$. Thus the channel $E_\Lambda$ sends an operator $a$ of the output observable algebra $\6A(J)$ to the operator 
\begin{equation}
E_\Lambda(a)=\1v_{[\Lambda,0^L]}^*\ a \ \1v_{[\Lambda,0^L]} 
\end{equation}
of the input algebra. We emphasize here that, in view of \cite{SchlWer00}, this is nothing else but the graph code which is associated with the subgraph $\Lambda^{IJ}_{IJ}$. 

We have already shown \cite[Corollay~VI.4]{Schl03} that each graph code can be implemented on a one-way quantum computer in a natural manner. For keeping the paper self-containt, we reformulate the corollary
here by adopting the notation introduced above: 

%%%
\begin{cor}
\label{encode}
The encoding channel $E_\Lambda$ fulfills the identity  
\begin{equation}
E_\Lambda=E_{[\11_I]} \ \alpha_{[\Lambda^{IJ}_{IJ}]} \ M_{[\11_I]} \ C_{[A_\Lambda]} 
\end{equation}
where the conditional phase space translation $C_{[A_\Lambda]}$ is associated with the classical channel $A_\Lambda\mathpunct:\9C(\Xi^J)\to\9C(\7F^I)$ which is given by 
\begin{equation}
\label{classical-device-1}
(A_\Lambda f)(p^I):=f(\Lambda^J_J\bar\Lambda^J_Ip^I,-\bar\Lambda^J_Ip^I) 
\end{equation}
for $f\in\9C(\Xi^J)$.
\end{cor}
%%%  

The statement of Corollary~\ref{encode} \cite[Corollay~VI.4]{Schl03} can be rephrased in terms of physical processes: (1) The output qudits are prepared in the shift invariant state. (2) One step of the discrete dynamics, associated with the subgraph $\Lambda^{IJ}_{IJ}$ is applied. (3) The input qudits are measured in $x$-basis. (4) Depending on the measurement outcome, a phase space translation is done. The relation between the measurement outcome and the phase space translation is described by the classical device $A_\Lambda$.  

%%%
\subsection{The decoding operation}
%%%
\label{impl-1-3}
Theorem~\ref{thm-1} tells us how an appropriate decoder $D_\Lambda$ can be constructed from the error correcting scheme $(\6w_\Lambda,\6v_\Lambda,\6u_\Lambda,\gamma_\Lambda)$. It is composed of the syndrome measurement operation and a local unitary correction operation which depends on the syndrome measurement outcome. 

%%%
\paragraph*{Syndrome measurement.}
Accordind to the identity (\ref{kraus-syndrome}), the Kraus operators for the syndrome measurement operation $S_{[\Lambda]}$ are given by the adjoints of the isometries in $\6v_\Lambda$. Thus the channel $S_{[\Lambda]}$ maps an operator valued function $a\mathpunct:\7F^L\to\6A(I)$ to the operator  
\begin{equation}
S_{[\Lambda]}(a)= \sum_{q^L\in\7F^L}\1v_{[\Lambda,q^L]}\ a(q^L) \ \1v_{[\Lambda,q^L]}^*
\end{equation}
of the output system $\6A(J)$. Concerning the concept of the one-way quantum computer, the syndrome measurement can be realized by four elementary operation as we prove in Appendix~\ref{app-3}: 

%%%
\begin{thm}\label{syndrome}
The channel $S_{[\Lambda]}$ fulfills the identity 
\begin{equation}
S_{[\Lambda]}=E_{[\11_{IL}]} \ \alpha_{[-\Lambda]} \  M_{[\11_J]} \  C_{[A'_\Lambda]} \  M_{[F_L]} 
\end{equation}
where the conditional phase space translation 
\begin{equation}
C_{[A'_\Lambda]}\mathpunct:\6A(IL)\to\9C(\7F^{JL},\6A(I))
\end{equation}
is associated with the classical channel $A'_\Lambda\mathpunct:\9C(\Xi^{IL})\to\9C(\7F^J)$ that is given by 
\begin{equation}
(A'_\Lambda f)(p^J):=f(0^{IL},\bar\Lambda^{IL}_Jp^J)
\end{equation}
for $f\in\9C(\Xi^{IL})$. 
\end{thm}
%%%

We interprete the statement of Theorem~\ref{syndrome} as follows: (1) The qudits at the input vertices $I$ and the syndrome vertices $L$ are prepared in the shift invariant state. (2) One elementary step of the dynamics with respect to the graph $-\Lambda$ is performed. (3) The output qudits are measured in $x$-basis. (4) The measurement outcomes of the output qudits are completely random and have to be compensated by a conditional phase space translation associated with the classical device $A'_\Lambda$. (5) The syndrome qudits are measured in $z$-basis.     

%%%
\paragraph*{The correction operation.} 
If the measurement result $q^L$ of the syndrome measurements is received, a classical program computes a solution $\xi^{IJ}=(p^{IJ},q^{IJ})$ of the system of equations 
\begin{equation}
\begin{array}{lcl}
\label{syndrome-table}
p^J-\Lambda^J_{IJL}q^{IJL}&=&0^J\\
&&\\
p^I-\Lambda^I_{J}q^{J}&=&0^I
\end{array}
\end{equation}
which is equivalent to the condition $\gamma_\Lambda(\xi^J,q^L,\xi^I)=0$. The phase space vector $\xi^I:=(p^I,q^I)$ is then the output of the classical device $A''_\Lambda\mathpunct:\9C(\Xi^I)\to\9C(\7F^L)$. By making use of the identities (\ref{corr-1}) and (\ref{corr-2}), the desired correction operation is just given by the conditional phase space translation 
\begin{equation}
C_{[A_\Lambda'']}\mathpunct:\6A(I)\to\9C(\7F^L,\6A(I))
\end{equation}
associated with the classical channel $A''_\Lambda$.

%%%
\paragraph*{The decoding operation.} 
Combining Theorem~\ref{thm-1} and Theorem~\ref{syndrome}, the decoding operation $D_\Lambda$ can be expressed in terms of five elementary operations
\begin{equation}\label{decode-2}
D_\Lambda=E_{[\11_{IL}]} \ \alpha_{[-\Lambda]} \ M_{[\11_J]} \ C_{[A'_\Lambda]} \ M_{[F_L]} \ C_{[A''_\Lambda]} \; .
\end{equation} 
The conditional phase space translations $C_{[A'_\Lambda]}$ and $C_{[A''_\Lambda]}$ can be preformed by one single conditional phase space translation. Namely, we combine the classical channels  $A''_\Lambda$ and $A'_\Lambda$ in the following manner: We build the classical device $B_\Lambda\mathpunct:\9C(\Xi^{I})\to\9C(\7F^{JL})$ according to
\begin{equation}\label{c-device}
(B_\Lambda f)(m^{JL})=f(p^I,q^I+\bar\Lambda^{I}_Jm^J)
\end{equation}
where the phase space vector $(p^I,q^I)$ is the unique solution of the equation (\ref{syndrome-table}) for $q^L=m^L-\bar\Lambda^{L}_Jm^J$ and $(p^J,q^J)\in\Xi_t^J$. We show in Appendix~\ref{app-2-1} that the following is true:

%%%
\begin{thm}\label{measure}
Using the notations, introduced above, the identity 
\begin{equation}
C_{[A_\Lambda']}\ M_{[F_L]}\ C_{[A''_\Lambda]}= M_{[F_L]}\ C_{[B_\Lambda]}
\end{equation}
holds.
\end{thm}
%%%

As a direct consequence of Theorem~\ref{measure} we obtain the corollary:

%%%
\begin{cor}\label{cor-decode}
The decoding operation $D_\Lambda$, associated with the error correcting scheme $(\6w_\Lambda,\6v_\Lambda,\6u_\Lambda,\gamma_\Lambda)$, fulfills the identity 
\begin{equation}
D_\Lambda=E_{[\11_{IL}]} \ \alpha_{[-\Lambda]}\ M_{[\11_J\otimes F_L]}\ C_{[B_\Lambda]}
\end{equation}
where the classical device $B_\Lambda$ is given by {\rm (\ref{c-device})}.
\end{cor}
%%%

Corollary~\ref{cor-decode} shows that, as the encoding operation, the decoder can be implemented on a one-way quantum computer by four elementary steps: (1) The qudits at the input vertices $I$ and the syndrome vertices $L$ are prepared in the shift invariant state. (2) One elementary step of the dynamics with respect to the graph $-\Lambda$ is performed. (3) The syndrome qudits are measured in $z$-basis and the output qudits are measured in $x$-basis. (4) Depending on the measurement outcomes of the output and syndrome qudits a phase space translation is performed, corresponding to the classical device $B_\Lambda$.

%%%
\subsubsection*{{\it Acknowledgment:}}
I am very grateful to Hans Briegel and Robert Raussendorf for supporting this investigation with many ideas. I would also like to acknowledge Reinhard Werner for his support and interesting and helpful discussions. 
This research project is funded by the ``Deutsche Forschungsgemeinschaft" which is also gratefully acknowledged. 
%%%

%%%
\begin{appendix}
%%%
%\begin{widetext}
%%%
\section{The proof of Theorem~\ref{thm-1}}
%%%
\label{app-1}
Each channel $T\in\9T$ has a Kraus representation of the form 
\begin{equation}
T(a)=\sum_{(q,x,y)} \bar t_{q,x} t_{q,y} \1w_x^* a\1w_y \; .
\end{equation}
Thus the composed channel $ETD$ has a Kraus representation by operators 
\begin{equation}
s_{(q,y,g,h)}=\sum_{x} t_{q,x} \sqrt{c(y,g,h)} \ \1u_h^*\1v_g^*\1w_x\1v_e \; .
\end{equation}
Let $(g(x),h(x))\in G\times H$ be a solution of the equation $\gamma(x,g(x),h(x))=0$ we find:
\begin{equation}
s_{(q,y,g,h)}=\sum_{x} t_{q,x} \sqrt{c(y,g,h)} \ \1u_h^*\1v_g^*\1v_{g(x)}\1u_{h(x)} \; .
\end{equation}
The isometries in $\6v$ are mutually orthogonal and we observe that
\begin{equation}
\1u_h^*\1v_g^*\1v_{g(x)}\1u_{h(x)}=\1u_h^*\1u_{h(x)}\delta_{g,g(x)} 
\end{equation}
holds. Now $\gamma(y,g,h)=0$ and $\gamma(x,g,h(x))=0$ imply that $h=h(x)$ and we conclude that all Kraus operators 
\begin{equation}
s_{(q,y,g,h)}=\sqrt{c(y,g,h)}\sum_{x} t_{q,x}\delta_{g,g(x)}\delta_{h,h(x)}
\11_{\9B(\2K)} 
\end{equation}
are multiples of the identity. Hence $ETD=\8{id}_{\9B(\2K)}$ follows. 
\hfill $\Box$
%%% 

%%%
\section{Proof of Theorem~\ref{e-c-scheme}}
%%%
\label{app-2}
%%%
The proof of Theorem~\ref{e-c-scheme} is prepared by showing three technical lemmas providing several useful relations.
%%%
\begin{lem}\label{com-rel}
Let $\Lambda$ be an admissible graph. Then the following is true:

\begin{enumerate}
\item
For a phase space vector $\xi^I=(p^I,q^I)\in\Xi^I$ and a register configuration $q^L\in\7F^L$ the identity 
\begin{equation}\label{id-1}
\1v_{[\Lambda,q^L]}\1w(\xi^I)=\tau(\Lambda,q^{IL})\1z(\Lambda^J_Iq^I)\1v_{[\Lambda,q^L]}\1z(p^I)
\end{equation}
holds. 

\item
For all phase space vectors $\xi^J=(p^J,q^J)\in\7F^J$ the identity 
\begin{multline}\label{id-2}
\1w(\xi^J)\1v_{[\Lambda,q^L]}=\tau(\Lambda,q^J-q^L)
\\
\times
\1z(p^J-\Lambda^J_Jq^J)\1v_{[\Lambda,q^L]}\1z(-\Lambda^I_Jq^J)
\end{multline}
holds.

\item 
For all register configurations $q^L\in\7F^L$ the identity 
\begin{equation}\label{id-3}
\1v_{[\Lambda,q^L]}=\1z(\Lambda^J_Lq^L)\1v_{[\Lambda,0^L]}
\end{equation}
holds.
\end{enumerate}
\end{lem}
%%%
\begin{proof}
Ad 1. We observe for a vector $\psi\in\H{I}$ and $\xi_1^I=(p^I_1,q_1^I)$ and $q^L$ that the  
identity 
\begin{multline}
[\1v_{[\Lambda,q^L]}\1w(\xi_1^I)\psi](q^J)\\
=\sqrt{d}^{|I|}\int\8dq^I\tau(\Lambda,q^{IJL}+q_1^I)\chi(p_1^I,q_I)\psi(q^I)
\end{multline}
is valid for all $q^J$. Now, we make use of the fact that the phases $\tau(\Lambda,\cdot)$ are a one-dimensional projective representation of $\7F^{IJL}$, i.e. $\tau(\Lambda,q^{IJL}+q_1^I)=\tau(\Lambda,q^{IJL})\chi(\Lambda^I_{JL}q^{JL},q_1^I)$ holds. This implies, by keeping in mind that there are no lines which connect syndrome (input) vertices, that the identity
\begin{multline}
\1v_{[\Lambda,q^L]}\1w(\xi_1^I)\psi
\\=\tau(\Lambda,q^L+q_1^I)\1z(\Lambda^J_Iq_1^I)\1v_{[\Lambda,q^L]}\1z(p_1^I)\psi
\end{multline}
is true.

\medskip
\noindent
Ad 2. For any vector $\psi\in\H{I}$ and register configurations $p^J_1,q_1^J$ and $q^L$ the identity (\ref{id-2}) is indeed true. Namely, if we apply the operator $\1w(\xi^J_1)\1v_{[\Lambda,q^L]}$ to a vector $\psi\in\H{I}$, then we find for a register configurations $q^J$: 
\begin{multline}
[\1w(\xi^J_1)\1v_{[\Lambda,q^L]}\psi](q^J)\\
=\sqrt{d}^{|I|}\chi(p_1^J,q^J)\int\8dq^I\tau(\Lambda,q^{IJL}-q_1^J)\psi(q^I)\; .
\end{multline}
We rewrite the product of phases $\chi(p_1^J,q^J)\tau(\Lambda,q^{IJL}-q_1^J)$ as a product of four phases 
according to 
\begin{multline}
\chi(p_1^J,q^J)\tau(\Lambda,q^{IJL}-q_1^J)=\tau(\Lambda,q_1^J-q^L)
\\\times
\chi(q^J,p_1^J-\Lambda^{J}_Jq_1^J)
\tau(\Lambda,q^{IJL})\chi(-\Lambda^{I}_Jq_1^J,q^I) \; .
\end{multline}
By means of this decomposition, the vector $\1w(\xi^J_1)\1v_{[\Lambda,q^L]}\psi$ can be obtained by the following sequence of operations: First we apply the multiplier operator 
$\psi\mapsto\psi_1:=\1z(-\Lambda^{I}_Jq_1^J)\psi$ which corresponds to multiplying the phases $\chi(-\Lambda^{I}_Jq_1^J,q^I)$ for all $q^I$. Then we apply the isometry 
$\psi_1\mapsto\psi_2:=\1v_{[\Lambda,q^L]}\psi_1$ which corresponds to multiplying the phases $\tau(\Lambda,q^{IJL})$ and integrating over the $q^I$ variables. In the next step, the multiplier $\psi_2\mapsto\psi_3:=\1z(p_1^J-\Lambda^{J}_Jq_1^J)\psi_2$ is performed due to multiplication by the phases $\chi(q^J,p_1^J-\Lambda^{J}_Jq_1^J)$. Finally, the constant phase $\tau(\Lambda,q_1^J-q^L)$ is left and we obtain the desired identity
\begin{multline}
\1w(\xi^J_1)\1v_{[\Lambda,q^L]}\psi=\tau(\Lambda,q_1^J-q^L)\psi_3
\\
=
\tau(\Lambda,q_1^J-q^L)
\1z(p_1^J-\Lambda^{J}_Jq_1^J)\1v_{[\Lambda,q^L]}\1z(-\Lambda^{I}_Jq_1^J)\psi \; .
\end{multline} 

\medskip
\noindent
Ad 3. In order to express the isometry $\1v_{[\Lambda,q^L]}$ in terms of the isometry $\1v_{[\Lambda,0^L]}$, we first consider $\1v_{[\Lambda,q^L]}\psi\in\H{J}$ at $q^J$: 
\begin{equation}
[\1v_{[\Lambda,q^L]}\psi](q^J)=\sqrt{d}^{|I|}\int\8dq^I\tau(\Lambda,q^{IJL})\psi(q^I) \; .
\end{equation}
The phase $\tau(\Lambda,q^{IJL})$ can be written as a product of three phases 
\begin{equation}
\tau(\Lambda,q^{IJL})=\chi(q^{J},\Lambda_L^{J}q^L)\tau(\Lambda,q^{IJ})\chi(q^{I},\Lambda_L^{I}q^L) \; .
\end{equation}
Multiplying the value $\psi(q^I)$ with $\chi(q^{I},\Lambda_L^{I}q^L)$ corresponds to an application of the multiplier operator $\1z(\Lambda_L^{I}q^L)$. Multiplying the phase $\tau(\Lambda,q^{IJ})$ and integrating the $q^I$ variables is nothing else but the applying the isometry $\1v_{[\Lambda,0^L]}$. Finally the multiplication by the phase $\chi(q^{J},\Lambda_L^{J}q^L)$ is an application of the multiplier operator $\1z(\Lambda_L^{J}q^L)$. Hence we get the desired result
\begin{equation}
\1v_{[\Lambda,q^L]}\psi=\1z(\Lambda_L^{J}q^L)\1v_{[\Lambda,0^L]}\psi
\end{equation}
and the identity (\ref{id-3}) follows.
\end{proof}
%%%

%%%
\begin{lem}\label{lem-a-2}
Let $\Lambda$ be an admissible graph with input vertices $I$, output vertices $J$, and syndrome vertices $L$. Then the identity 
\begin{equation}\label{equ-1}
\1w_{[\Lambda,\xi^J]}\1v_{[\Lambda,0^L]}=\1v_{[\Lambda,q^L]}\1u_{[\Lambda,\xi^I]}
\end{equation}
is valid if $\gamma(\xi^J,q^L,\xi^I)=0$ holds.
\end{lem}
%%%
\begin{proof}
According to Lemma~\ref{com-rel}, we conclude from (\ref{id-1}) that  
\begin{multline}
\1w_{[\Lambda,\xi^J]}\1v_{[\Lambda,0^L]}=\tau(-\Lambda,q^J)\1w(\xi^J)\1v_{[\Lambda,0^L]}
\\
=\1z(p^J-\Lambda^J_{J}q^{J})\1v_{[\Lambda,0^L]}\1z(\Lambda^I_Jq^J)
\end{multline}
holds for all phase space vectors $\xi^J=(p^J,q^J)$. By (\ref{id-3}), we express the isometry $\1v_{[\Lambda,0^L]}$ in terms of $\1v_{[\Lambda,q^L]}$ which yields 
\begin{multline}
\1w_{[\Lambda,\xi^J]}\1v_{[\Lambda,0^L]}
=\1z(p^J-\Lambda^J_{J}q^{J}-\Lambda_L^{J}q^L)\1v_{[\Lambda,q^L]}
\\\times\1z(\Lambda^I_Jq^J)
\end{multline}
where we have used the assumption that $\Lambda^I_L=0$. Applying the identity (\ref{id-1}) implies that 
\begin{multline}
\1w_{[\Lambda,\xi^J]}\1v_{[\Lambda,0^L]}
=\1z(p^J-\Lambda^J_{IJL}q^{IJL})\1v_{[\Lambda,q^L]}
\\\times\1w(\Lambda^I_Jq^J,q^I)
\end{multline} 
is true. Now, the condition $\gamma_\Lambda(\xi^J,q^L,\xi^I)=0$  implies (\ref{equ-1}) since then $p^J-\Lambda^J_{IJL}q^{IJL}=0^J$ and $p^I=\Lambda^I_Jq^J$ are valid. 
\end{proof}
%%%

%%%

%%%
\begin{lem}\label{lem-a-1}
Let $\Lambda$ be an admissible graph which is associated with a $t$-error correcting code, then the following is true:
\begin{enumerate} 
\item
For each $\xi^J\in\Xi^J$ there exist $q^L\in\7F^L$ and $\xi^I\in\Xi^I$ such that $\gamma_\Lambda(\xi^J,q^L,\xi^I)=0$. 
\item
If the identity 
\begin{equation}
\gamma_\Lambda(\xi_1^J,q^L,\xi_1^I)=\gamma_\Lambda(\xi_2^J,q^L,\xi_2^I)=0
\end{equation}
hold for $\xi_1^J,\xi_2^J\in\Xi_t^J$ and $q^L\in\7F^d$, then $\xi_1^I=\xi_2^I$ follows.
\end{enumerate}
\end{lem}
%%%
\begin{proof}
Ad 1.: We first observe that, according to our assumptions, the block matrix $\Lambda^J_{IL}$ has an inverse $\bar\Lambda^{IL}_J$. Then the system of equations (\ref{gam}) is equivalent to 
\begin{equation}
\begin{array}{lcl}
q^{IL}&=&\bar{\Lambda}_J^{IL}[p^J-\Lambda^J_{J}q^{J}]\\
&&\\
p^I&=&\Lambda^I_{J}q^{J}
\end{array}\; .
\end{equation} 
\medskip

\noindent
Ad 2.: Suppose that the identities $\gamma(\xi_i^J,q^L,\xi_i^I)=0$, $i=1,2$, hold, then we conclude that 
\begin{equation}\label{conclude-1}
\begin{array}{llcl}
&p_1^J-p^2_J-\Lambda^J_{IJ}(q_1^{IJ}-q_2^{IJ})&=&0\\
\text{and}&&&\\
&p_1^I-p_2^I-\Lambda^I_{J}(q_1^J-q_2^J)&=&0
\end{array}
\end{equation} 
is valid. If $\xi_1^J,\xi_2^J\in\Xi_t^J$ have weight smaller or equal than $t$, then there exists a subset $E\subset J$ with at most $2t$ elements with $\xi_i^J=\xi_i^E$, $i=1,2$. In this case, (\ref{conclude-1}) is equivalent to 
\begin{equation}\label{conclude-2}
\begin{array}{llcl}
&p_1^E-p_2^E-\Lambda^E_{IE}(q_1^{IE}-q_2^{IE})&=&0\\
\text{and}&&&\\
&\Lambda^{J\setminus E}_{IE}(q_1^{IE}-q_2^{IE})&=&0\\
\text{and}&&&\\
&p_1^I-p_2^I-\Lambda^I_{E}(q_1^E-q_2^E)&=&0
\end{array}
\end{equation} 
Since $\Lambda$ is associated with a $t$-error correcting code, $q_1^I-q_2^I=0$ and $\Lambda^I_{E}(q_1^E-q_2^E)=0$ follows. Thus we obtain from (\ref{conclude-2}) that $p^I_1=p^I_2$, i.e.
$\xi^I_1=\xi^I_2$. 
\end{proof}
%%%

\begin{proof}[Proof of Theorem~\ref{e-c-scheme}]
%%%
As described in Subsection~\ref{stab-codes-2}, we construct from an admissible graph $\Lambda$  a unitary error basis 
$\6w_\Lambda$, a complete set of mutually orthogonal isometries $\6v_\Lambda$, a set of unitary correction operations  $\6u_\Lambda$, and a candidate for syndrome table $\gamma_\Lambda$. 

According to Lemma~\ref{lem-a-2} and Lemma~\ref{lem-a-1}, the four-tuple $(\6w_\Lambda,\6v_\Lambda,\6u_\Lambda,\gamma_\Lambda)$ satisfies the defining properties of an error correcting scheme which have been introduced in Subsection~\ref{sec-1-1}.  
\end{proof}
%%%

%%%
\section{Proof of Theorem~\ref{syndrome}}
%%%
\label{app-3}
Let $\Lambda$ be an admissible graph with input vertices $I$, output vertices $J$, and syndrome vertices $L$. Since the block matrix $\Lambda_J^{IL}$ is invertible, the weighted graph $-\Lambda$ can be regarded as ``basic graph"\footnote{A basic graph $\Gamma$ lives on input vertices $I$ output vertices $J$ and measuring vertices $K$. Its defining condition is that the block matrix $\Gamma^{JK}_{IK}$ has maximal rank.} with input vertices $J$ and output vertices $IL$ and an empty set of measuring vertices. According to \cite{Schl03}, the operation 
\begin{equation}
S'_{[\Lambda]}:=E_{[\11_{IL}]}\ \alpha_{[-\Lambda]}\ M_{[\11_J]}\ C_{[A'_\Lambda]}
\end{equation}
is implemented by a single unitary Kraus operator $\1u_{[-\Lambda,J]}$ which acts on $\H{J}$ according to 
\begin{multline}
(\1u_{[-\Lambda,J]}\psi)(q^{IL})\\=\sqrt{d}^{|IL|}\int\8dq^J\tau(-\Lambda,q^{IJL})\psi(q^J)
\end{multline}
The Kraus representation for the measurement $M_{[F_L]}$ of the syndrome qudits is given in terms of the co-isometries $\Phi_L^*\1z(q^L)^*F_L^*=\Phi_L^*F_L^*\1x(q^L)^*$, $q^L\in\7F^L$. One easily observes for a vector $\psi\in\H{IL}$ that $(\Phi_L^*F_L^*\1x(q^L)\psi)(q^I)=\sqrt{d}^{-|L|}\psi(q^{IL})$ holds for all $q^{I}\in\7F^I$. As a consequence we find
\begin{multline}
(\Phi_L^*F_L^*\1x(-q^L)^*\1u_{[-\Lambda,J]}\psi)(q^{I})\\=\sqrt{d}^{|I|}\int\8dq^J\tau(-\Lambda,q^{IJL})\psi(q^J)
\end{multline}
which implies $\Phi_L^*F_L^*\1x(q^L)^*\1u_{[-\Lambda,J]}=\1v_{[\Lambda,q^L]}^*$. The syndrome measurement is implemented by the co-isometries $\1v_{[\Lambda,q^L]}^*$, $q^L\in\7F^L$, and we conclude that the desired identity
\begin{equation}
S_{[\Lambda]}=E_{[\11_{IL}]}\ \alpha_{[-\Lambda]}\  M_{[\11_J]}\  C_{[A'_\Lambda]}\ M_{[F_L]} 
\end{equation}
is satisfied. \hfill $\Box$
%%%

%%%
\section{Proof of Theorem~\ref{measure}}
%%%
\label{app-2-1}
We first observe that the channel $C_{[A_\Lambda']}\ M_{[F_L]}\ C_{[A''_\Lambda]}$, 
which maps observables of the input system $\6A(I)$ to $\6A(IL)$-valued functions on $\7F^J$, has a Kraus representation by   
the operators  
\begin{equation}
r_{[m^J,q^L]}:=\1w(\xi^I_{[q^L]})^*\Phi_L^*F_L^*\1x(q^L)^*\1x(\bar\Lambda^{IL}_J m^J)^*
\end{equation}
where $\xi^I_{[q^L]}$ is defined by the syndrome table $\gamma_\Lambda$: If there exists a phase space vectors $\xi^J\in\Xi^J_t$ and $\xi^I$ with $\gamma_\Lambda(\xi^J,q^L,\xi^I)=0$, then we define $\xi^I_{[q^L]}:=\xi^I$. Since $\gamma_\Lambda$ is a syndrome table, the vector $\xi^I_{[q^L]}=\xi^I$ is uniquely determined by the error syndrome $q^L$. If there is no phase space vector $\xi^J\in\Xi_t^J$ with $\gamma_\Lambda(\xi^J,q^L,\xi^I)=0$, then we put $\xi^I_{[q^L]}:=0^I$. 

On the other hand, the channel $M_{[F_L]}\ C_{[B_\Lambda]}$ has a Kraus representation by operators 
\begin{equation}
r'_{[m^J,m^L]}=\Phi_L^*F_L^*\1x(m^L)\1w(p^I,q^I+\bar\Lambda^{I}_Jm^J)^*
\end{equation}
where the phase space vector $\xi^I=(p^I,q^I)$ is given by $\xi^I=\xi^I_{[m^L-\bar\Lambda^{L}_Jm^J]}$. The Weyl operator $\1w(\xi^I)^*$ commutes with the co-isometry $\Phi_L^*F_L^*$ which implies that the identity 
$r_{[m^J,q^L]}=r'_{[m^J,m^L]}$ holds for $q^L=m^L-\bar\Lambda^{L}_Jm^J$. This implies the theorem.
\hfill $\Box$
%%%

%%%
\end{appendix}
%%%
\input{decode20030605-ref.tex}
%%%
\end{document}

%% file: decode20030605-ref.tex
%%%

%%%